\documentclass[prl,twocolumn,superscriptaddress,9pts]{revtex4}

\newcommand{\Eqref}[1]{(\ref{#1})}

\newcommand{\ket}[1] {\mbox{$ \vert #1 \rangle $}}

\newcommand{\bra}[1] {\mbox{$ \langle #1 \vert $}}

\newcommand{\abs}[1] {\mbox{$ \vert #1 \vert $}}

\usepackage{color}      

\usepackage{epsfig}

\newcommand{\ba}{\begin{eqnarray}}

\newcommand{\ea}{\end{eqnarray}}

\newcommand{\sba}{\begin{subeqnarray}}

\newcommand{\sea}{\end{subeqnarray}}

\begin{document}

\vskip 1truecm

 \title{Inflationary spectra and violations of Bell inequalities}
 \vskip 1truecm
 \author{David Campo}
 \affiliation{Department of Applied Mathematics, University of Waterloo,
 University Avenue, Waterloo, Ontario, N2L 3G1 Canada}
 \author{Renaud Parentani}
 \affiliation{Laboratoire de Physique Th\'{e}orique, CNRS UMR 8627,
 B\^atiment 210, Universit\'{e} Paris XI, 91405 Orsay Cedex, France}
 \begin{abstract}
 In spite of the macroscopic character of the primordial fluctuations,
 the standard inflationary distribution (that obtained using linear mode
 equations) exhibits inherently quantum properties, that is, properties
 which cannot be mimicked by any stochastic distribution.
 This is demonstrated by a Gedanken experiment
 for which certain Bell inequalities are violated.
 These violations are
 {\it in principle} measurable
 because, unlike for Hawking radiation from black
 holes, in inflationary cosmology we can have access to both members
 of correlated pairs of modes delivered in the same state.
 We then compute the effect of decoherence and
 show that the violations persist provided the decoherence
 level (and thus the entropy) lies below
 a certain non-vanishing threshold. Moreover, there exists
 a higher threshold above which no violation of any Bell
 inequality can occur. In this regime, the distributions are ``separable''
 and can be interpreted as stochastic ensembles of fluctuations.
 Unfortunately, the precision which is required to have access to
 the quantum properties
 is so high that, {\it in practice},  an observational verification
 seems excluded.
 \end{abstract}
 \maketitle

 The inflationary paradigm \cite{inflation0} successfully accounts
 for the properties of primordial spectra revealed
 by the combined analysis of CMBR temperature anisotropy and
 Large Scale Structure spectra \cite{WMAP}.
 In particular, it predicts
 that the distribution of primordial fluctuations is
 homogeneous, isotropic and Gaussian,
 and that the power spectrum is nearly scale invariant (simply because the
 Hubble radius was slowly varying during inflation).

 Surprisingly, inflation implies that
 density fluctuations arise from the
 amplification of vacuum fluctuations \cite{inflation};
 because of backreaction effects,
 the vacuum is indeed the only possible initial state
 \cite{RenaudCMB}.
 In addition of being amplified, the
 modes of opposite wave-vectors $\bf k$ and $-\bf k$
 end up highly correlated. More precisely, using linear
 mode equations, the vacuum evolves into
 a product of two-mode squeezed states
 \cite{squeezing,squeezing2,CaveSchumaker,AllenFlanagan}.
 The highly squeezed character of the distribution
 implies the vanishing of the variance in one direction in phase space. 
 This direction is that of the decaying mode \cite{squeezing2}.
 The observational consequence
 of this squeezing are the acoustic peaks in the
 temperature anisotropy
 spectrum \cite{Albrecht,Staroentropy}.

 In spite of the macroscopic character of the mode amplitudes,
 we shall show that the inflationary 
 distribution
 is still entangled in a quantum mechanical sense.
 To prove this, we shall provide observables
 able to distinguish quantum correlations from stochastic correlations.
 At this point, it is important to notice that,
 unlike for Hawking radiation from black holes,
 we have {\it in principle} access to
 the purity of the state since,
 both members of two-mode sectors in the same state
 can be simultaneously
 observed on the last scattering surface \cite{CP2}.

 Another important element should now be discussed:
 the linear mode equation is only approximate.
 Indeed, even in the simplest inflationary models there exists
 gravitational interactions
 which couple sectors with different $\bf k$'s, and induce
 non-Gaussianities \cite{Maldacena03}.
 However, as in the $BCS$ description of super-conductivity \cite{BCS},
 the weakness of the interactions allows to approximate
 the distribution by a product of
 Gaussian two-mode distributions \cite{Staroentropy,CPfat}.
 The non-linearities will then affect the power spectrum
 as if some decoherence effectively occurred.
 In this sense, inflationary distributions
 belong to the class of Gaussian homogeneous distributions
 obtained by slightly decohering the standard distribution
 derived with linear mode equations.
 Notice also that in general, we have an experimental access
 to the state of a system only through a
 truncated hierarchy of it's Green functions,
 the Gaussian ansatz being
 the lowest order (Hartree) approximation.

 In the absence of a clear evaluation of the importance of
 non-linearities \footnote{{\it Added Note:}
 The question of the importance of decoherence effects
 induced by the weak non-linearities neglected in the
 standard treatment has been recently
 addressed in a couple of preprints \cite{morebs}.
 The non-linearities have been treated in the
 Gaussian approximation, as in
 \cite{Zu,E}. Therefore the reduced density matrices
 belong to the class of
 partially decohered matrices described in Eqs. (\ref{momenta}) and
 (\ref{momenta2}), and considered in more details in \cite{CPfat}.
 To simplify the calculation,
 the environement has been effectively described by local correlation
 functions, i.e. by only short wave length modes.
 Since this simplification still requires to be legitimized,
 the decoherence level at the end of inflation, i.e. the value of $\delta$,
 is still unknown.\label{foot1}},
 it is of value to
 phenomenologically analyze the above class.
 It is characterized by three $k$-dependent parameters.
 The first governs the power, see $n_k$ in \Eqref{momenta}.
 The second gives the orientation of the squeezed direction in phase space,
 whereas the third controls the strength of the correlations between modes
 with opposite momenta. The latter is strongly affected by
 decoherence effects, and shall be used to parameterize the decoherence
 level.
 It has been understood \cite{Albrecht,Staroentropy}
 that this level cannot be too high so as to preserve
 the well defined character of the acoustic peaks.
 However what is lacking in the literature
 concerning the quantum-to-classical transition is
 an operational identification of the subset of
 distributions exhibiting quantum correlations.

 To fill the gap, we propose a Gedanken experiment
 which shows that certain Bell inequalities are violated
 when using the standard distribution.
 We then show that the violation persists provided that the
 decoherence level lies below a certain threshold.
 Finally we point that there exists a higher threshold above which
 no violation of any Bell inequality can occur.
 The corresponding distributions are {\it separable}
 (see below for the definition)
 and can be interpreted as stochastic ensembles.

 In inflationary models based on one inflaton field,
 the linear metric (scalar and tensor) perturbations
 around the homogeneous background
 are governed by massless minimally coupled scalar fields
 obeying canonical commutation relations \cite{MukhaPhysRep}.
 The scalar metric perturbations are driven by
 the inflaton fluctuations and correspond to perturbations along
 the background trajectory, called
 adiabatic perturbations \cite{Wands}.
 At the end of inflation, the homogeneous inflaton condensate decays
 and heats up matter fields.
 After inflation, during the radiation dominated era,
 the adiabatic perturbations correspond to
 density perturbations of the matter fields
 (radiation, dark matter, ...) which all start to oscillate in phase.
 The fluctuations orthogonal to these, called iso-curvature,
 are not excited on cosmological scales in one inflaton field models.
 Therefore, in the linear approximation, the phase and
 amplitude of the ${\bf k}$-th Fourier mode
 of each matter density fluctuation is related,
 via a time dependent transfer matrix,
 to the value of  $\phi_{\bf k}$ and its time
 derivative evaluated at the end of inflation
 ($\phi$ being the canonical
 field governing scalar metric fluctuations during inflation).
 This implies that the properties of the 
 correlations of the density fluctuations
 are the {\it same} as those of $\phi_{\bf k}$.

 We now briefly outline how one obtains highly squeezed two-mode
 states \cite{squeezing,squeezing2}.
 During inflation, in the linearized treatment,
 each $\phi_{\bf k}$ evolves under its own
 Hamiltonian
 \ba \label{Hquad}
   H_{\bf k} =   \frac{1}{2}   \, \left[
   \vert  \partial_{\eta}  \phi_{\bf k} \vert^2
   + \left(k^2 - \frac{\partial_{\eta}^2 a}{a} \right) \,
   \vert   \phi_{\bf k} \vert^2
   \right]
   \, ,
 \ea
 where $\eta$ is the conformal time $d\eta = dt /a$ and $a$ is the scale
 factor.
 To follow the mode evolution after the reheating time $\eta_r$,
 we continuously extend
 the inflationary law to a radiation dominated phase wherein
 $a \propto \eta$.
 In quantum settings,
 the initial state of the relevant modes (i.e. today observable
 in the CMBR)
 is fixed by the kinematics of inflation \cite{RenaudCMB}:
 these were in their ground
 state about $70$ e-folds before the end of inflation
 (the minimal duration of inflation to include
 today's Hubble scale inside a causal patch).
 From horizon crossing $k /a = H$ till the reheating time,
 $(k^2 - \partial_{\eta}^2 a/a)$ in \Eqref{Hquad}
 is negative.
 As a result, at the end of inflation, the initial vacuum has
 evolved into a
 tensor product of highly squeezed two-mode states.

 The resulting distribution
 belongs to the class of Gaussian homogeneous
 distributions, see \cite{CPfat} for more details. These
 are characterized by their two-point functions,
 best expressed as
 \ba \label{momenta}
   \langle \hat a_{\bf k}^{\dagger} \,   \hat  a_{\bf k'} \rangle
   = n_k  \, \delta^3({\bf k} - {\bf k}' ) \, , \quad
   \langle \hat a_{\bf k}  \, \hat a_{\bf k'} \rangle
   = c_k \, \delta^3({\bf k} + {\bf k}' )
   \, .
 \ea
 The destruction operator $\hat a_{\bf k}$ is defined
 by $\hat a_{\bf k}\, e^{-ik\eta_r} =
 \sqrt{k/2}(\hat\phi_{\bf k} + i \partial_\eta \hat\phi_{\bf k}/k)$
 where $\hat\phi_{\bf k}$ is evaluated at $\eta_r$.
 The mean occupation number governs the power spectrum,
 as shall be explained after Eq. \Eqref{momenta2}.
 To meet the observed r.m.s. 
 amplitude of the order of $10^{-5}$,
 one needs $n_k \sim 10^{100}$, i.e. highly excited states.
 The phase $\arg(c_k)$ gives the
 orientation of the squeezed
 direction in phase space at $\eta_r$.
 In inflation, using the above phase conventions, 
 one gets $\arg(c_k) =\pi + O(n_k^{-3/4})$.
 Finally, the norm of ${c_k}$ governs the strength of the
 correlations between partner modes
 ${\bf k}, {-\bf k}$, i.e.,
 the level of the coherence of the distribution.
 To parameterize the (de)coherence level,
 we shall work at fixed $n$ and $\arg(c)$
 (in the sequel we drop the $k$ indexes),
 and write the norm $\abs{c}$ as
 \ba
   \abs{c}^2 = (n+1)(n-\delta) \, .
 \ea
 The standard distribution obtained in the linear treatment
 is maximally coherent and corresponds to $\delta = 0$.
 The least coherent distribution,
 a product of two thermal density matrices, corresponds
 to $\delta = n$.

 The physical meaning of $\delta$ is revealed by
 decomposing the adiabatic modes 
 in terms of the amplitudes ($g, d$) of
 the growing and decaying solutions.
 Taking into account the time dependence of the corresponding
 transfer matrix, {\it any} matter density fluctuation can be used.
 For simplicity, we shall use the massless field $\phi$
 extended in the radiation dominated era.
 In this case, the transfer matrix
 of $\hat a_{\bf k}$ is simply $e^{-i k \eta}$.
 Decomposing
 \ba
   \hat \phi_{\bf k}(\eta) =
   \hat g_{\bf k} \frac{\sin (k\eta)}{\sqrt{k}} +
   \hat d_{\bf k} \frac{\cos (k\eta)}{\sqrt{k}} \, ,
 \ea
 Eqs. \Eqref{momenta} give
 \ba   \label{momenta2}
   \langle \hat g_{\bf k} \hat g_{\bf k}^{\dagger}
   \rangle &=& n + \frac{1}{2} - Re(c) =
   2n \, \left( 1 + O(\frac{\delta}{n})\right)
   \, , \nonumber \\
   \langle \hat d_{\bf k} \hat d_{\bf k}^{\dagger}
   \rangle &=&  n + \frac{1}{2} + Re(c)
   = \frac{\delta}{2} + O(n^{-1/2})
   \, .
 \ea
 The last expression in each line is valid when
 the decoherence is weak, i.e. $\delta \ll n$.
 In this regime, the power spectrum
 $P_k = k^3 \langle \hat \phi_{\bf k}(\eta)\hat \phi_{-{\bf k}}(\eta)
 \rangle \simeq k^2 n_k \sin^2(k \eta)$ is dominated
 by the growing mode. At fixed $\eta$, it
 therefore displays peaks and zeros as $k$ varies.
 From the last equation \Eqref{momenta2},
 one sees that the decoherence
 level $\delta$ fixes the power of the decaying mode.
 (The same conclusions would have been reached
 had we considered dark matter or temperature perturbations.)

 Even though Eqs. \Eqref{momenta}
 univocally determine the corresponding
 (Gaussian)
 distribution, they are unable to
 sort out the distributions possessing quantum
 properties from those which have
 lost them, or in other words, to determine the ranges of $\delta$
 characterizing these two classes. To operationally do so, it is necessary
 to introduce operators which are not polynomial in
  $\hat g_{\bf k}$ and $\hat d_{\bf k}$ 
 \footnote{Indeed
 the expectation values of
 Weyl ordered products of the field amplitude
 and its conjugate momentum
 (or equivalently $g$ and $d$)
 behave as in classical statistical mechanics
 when the Wigner function is positive,
 as is always the case for Gaussian states}.

 In what follows, we shall use operators based on coherent states.
 These obey
 $\hat a_{\bf k} \ket{v, {\bf k}} = v \ket{v, {\bf k}} $
 and
 $\hat a_{\bf - k} \ket{w, {\bf -k}} = w \ket{w, {\bf -k}} $.
 They are minimal uncertainty states and each of them
 can be considered as the quantum counterpart of a point in
 phase space, here a classical fluctuation with definite phase and
 amplitude.
 This correspondence is excellent in the regime
 $n \gg 1$. Moreover, they play a
 key role when considering decoherence:
 when modes are weakly coupled to an environment, the reduced
 density matrix becomes diagonal in the basis of coherent states \cite{Zu},
 or other minimal uncertainty states \cite{E}.

 Coherent states are particularly useful in our context
 because they
 will allow us to sort out entangled quantum distributions
 from stochastic ones. The reason is that coherent states
 can probe the detailed properties of the distribution.
 In particular, the probability to find a
 particular
 classical fluctuation is given by the expectation value
 of the projector on the corresponding (two-mode) coherent state,
 namely
 \ba \label{proj}
   \Pi(v,w) = \ket{v, {\bf k}}\bra{v, {\bf k}} \otimes
   \ket{w, {\bf -k}}\bra{w, -{\bf k}} \, .
 \ea
 The probability is
 \ba \label{proba}
   Q(v,w; \delta) &=&  {\rm Tr}[\rho_2(\delta) \, \Pi(v,w) ]
   \, , \nonumber \\
   &=& \frac{1}{n+1} \exp\left[ - \frac{\abs{v}^2}{(n+1)} \right]
   \times
    \nonumber\\
   && \quad
    \frac{1}{1+ \delta} \exp\left[ -
   \frac{\abs{w - \bar w(v)}^2}{1+\delta}  \right]
   \, , \quad
 \ea
 where $\rho_2(\delta)$ is the matrix density of the two-mode system. 
 We have written $Q(v,w; \delta)$ in an asymmetric form to make
 explicit the power of the growing mode ($=n+1$),
 and the much smaller width
 ($={1+\delta}$) governing the dispersion of the values of $w$ around
 $\bar w(v) =  v^* c/(n+1)$, the {\it conditional}
 amplitude of the partner mode, given $v$. 
 Had we used a projector on a one-mode coherent state,
 we would have gotten only the first Gaussian.
 In fact, as we shall see,
 to have access to
 the (residual) quantum properties
 of the distribution,
 one must use the two-mode projectors \Eqref{proj}.
 As explained in \cite{CP2},
 these projectors also allow to compute conditional
 values which cannot be expressed in terms of mean values. For instance,
 ${\rm Tr}[\Pi \rho \Pi \, \hat \phi(\eta,{\bf x})]$ gives
 the space-time pattern of fluctuations when the set of
 configurations specified by the projector
 $\Pi$ is realized.

 Given the macroscopic character of mode amplitudes in inflationary
 cosmology, it is remarkable that the projectors \Eqref{proj}
 can violate Bell inequalities.
 To understand the origin of this possibility,
 it is necessary to define the
 class of {\it separable} states \cite{Werner}.
 A two-mode state is said separable if
 it can be written as a
 positive sum of products of one-mode density matrices
 Separable Gaussian states can
 all be written in terms of the projectors \Eqref{proj} as \cite{CPfat}
 \ba \label{separable}
    \rho_2^{\rm sep.}(\delta) =
    \int\!\!\frac{d^2v}{\pi} \frac{d^2w}{\pi} \,
    P(v,w;\delta) \, \Pi(v,w) \, .
 \ea
 The function $P$ is given by 
 \ba
   P(v,w;\delta) &=&
   \frac{1}{\Delta'} \exp\left[-\frac{\abs{v}^2}{n}\right] \times
   \exp \left[ - \frac{\abs{w - \tilde w}^2}{\Delta'/n }  \right] \, ,
   \quad
 \ea
 with $\tilde w=c v^*/n$ and
 $\Delta' = n^2 - \abs{c}^2 \geq 0$.
 The latter implies $\abs{c} \leq n$, or $\delta \geq n/(n+1) \simeq 1$ for
 $n \gg 1$.
 (The limiting case $\abs{c}= n$, $\delta = n/(n+1)$  is interesting:
 the second exponential becomes
 a double Dirac delta which enforces
 $w= c v^*/n = - v^*$ in phase and amplitude. In other words, 
 for each {\it two-mode} sector, there is only {one} 
 fluctuating quantity,
 since the second mode is completely fixed by its partner.
 In inflationary cosmology,
 the corresponding density matrix
 can be viewed as the quantum analogue
 of the usual stochastic distribution
 of growing modes.
 Indeed, the entropy of this quantum distribution is $\ln(n)$ per two-mode,
 and this is the entropy
 of the stochastic distribution for each 
 growing mode \cite{CPfat}.
 This quantum-to-classical correspondence is
 corroborated by the fact that off-diagonal matrix elements of
 $\rho(\delta)$
 in the coherent state basis vanish
 precisely when $\delta  > 1$.

 The physical meaning of separable states comes from the fact that
 all states of the form \Eqref{separable} can be obtained by
 the following classical protocol \cite{Werner}: when a random generator
 produces the four real numbers encoded in $(v,w)$
 with probability $P$, two space-like separated
 observers performing separate
 measurements on the subsystems
 $\bf k$ and $-{\bf k}$ respectively, prepare them
 into the two-mode coherent state $\ket{v}\ket{w}$.
 Non-separable states
 can only be produced by letting the two parts of the system interact.
 Only these are quantum mechanically 
 entangled.

 By construction, the statistical properties of separable
 states can be interpreted classically.
 In particular, they cannot violate Bell inequalities \cite{Werner}.
 In what follows we shall study 
 the ``Clauser-Horne'' inequality  \cite{CH,Wodkievicz98} because
 it is based on $Q$ of \Eqref{proba}. It reads
 \ba \label{Bell}
   {\cal C}(v,w;\delta) &=& [ Q(0,0;\delta) + Q(v,0;\delta)
   \\
   && + \, Q(0,w;\delta) - Q(v,w;\delta)] \times
   \left(\frac{n+1}{2} \right) \leq  1 \, .
  \nonumber
 \ea
 We can now search for distributions, i.e. values of $\delta$,
 and for
 configurations $v$ and $w$ which maximize ${\cal C}$.
 The maximization with respect to $w$
 gives $\arg(c^* v w) = \pi$ and $\abs{w}= \abs{v}$.
 We fix
 the arbitrary phase of $v$
 by $2 \arg(v)= \arg(c)$, so that ${\cal C}$ is maximum along the 'line'
 $w=-v$.
 In Fig. 1 we have plotted ${\cal C}(v,-v,\delta)$
 for three values of $\delta$.
 \begin{figure}[h]
    \resizebox{8cm}{4cm}{\includegraphics{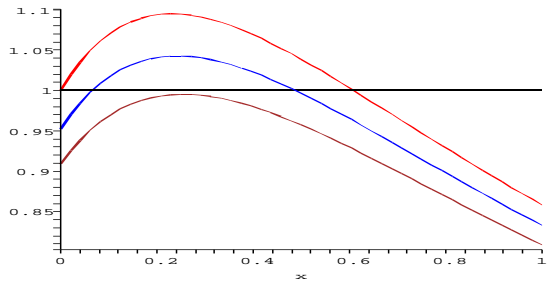}}
 \caption{{\it The loss of violation as decoherence increases.}
  We have represented ${\cal C}(v,-v,\delta)$ as a function of $x=\abs{v}^2$
 for $n=100$ and for three values of $\delta$:
 $0$ (upper), $0.05$ (middle), and
 $0.1$ (lower).
 The horizontal line (${\cal C}=1$)
 is the maximal value allowed
 by classically correlated states.}
\end{figure}
 The maximum with respect to the norm of $v$
 is reached for
 \ba
   \frac{\abs{v_M(\delta)}^2}{1+ \delta} &=&
   \frac{\ln \left[ 1 + \sqrt\frac{n-\delta}{n+1} \right]}{1 +
   2 \sqrt\frac{n-\delta}{n+1}}
   = \frac{\ln 2}{3}
   \left[ 1 + O(\delta/n)   \right] \, . \quad
 \ea
 The maximal value is
 \ba \label{vi}
  {\cal C}_M(\delta) &=&
  \frac{1}{2(1+\delta)} \times  \left[ 1 + \frac{3}{2^{4/3}}
  + O\left(\frac{1 + \delta}{n}\right)
  \right] \,  .\quad \quad
 \ea
 The inequality \Eqref{Bell} is thus violated for
 \ba  \label{range}
   \delta < \frac{-1+3/2^{4/3}}{2} \simeq 0.095 \, ,
 \ea
 irrespectively of the value of $n$ when $n \gg 1$.

 From the last two equations we learn that
 {\it Bell inequality \Eqref{Bell} is violated
 by the standard inflationary distribution} ($\delta =0$).
 Notice that this violation is maximal, as one might have expected,
 since the two-mode correlations are the strongest in this state.
 More importantly, if $\delta$ obeys \Eqref{range},
 the violation persists
 in the regime of highly amplified modes obtained
 in inflationary cosmology.

 In conclusion, the principle results
 of this Rapid Communication
 are the following.
 First, in spite of the macroscopic character of adiabatic fluctuations,
 the standard inflationary distribution possesses quantum features
 which cannot be mimicked by any stochastic distribution. 
 Second, these features are operationally revealed
 by a well defined procedure based on the
 violation of
 the Bell inequality (\ref{Bell}).
 Third, the projectors used in this inequality have a clear
 meaning in cosmology:
 they give the probability that a particular semi-classical fluctuation
 be realized.
 Fourth, the mere existence of decoherence effects
 is not sufficient to eliminate the quantum properties.
 To do so, decoherence should be strong enough
 so as to induce $\delta \geq 1$,
 that is, so that the distribution becomes separable.

 The threshold value $\delta = 1$ therefore plays a double
 role.
 First, as previously noticed, the distribution with $\delta = 1$
 possesses an entropy ($=\ln n$ per two-mode)
 which is equal to that of the classical distribution
 of growing modes.
 Second, separability is
 the condition for
 distinguishing quantum from classical distributions,
 see e.g. \cite{Halliwell} where
 it was used to define the time of decoherence.
 To our knowledge, besides the present work, this
 criterion
 of the study of the quantum-to-classical transition
 has not been used in inflationary cosmology.

 Let us now briefly address two additional questions.
 Firstly, to what extend 
 the violation of the inequality \Eqref{Bell} is verifiable ?
 We start by pointing out that there is no physical principle which
 prevents
 evaluating the four terms in Eq. \Eqref{Bell}.
 Because of isotropy, in a given comoving volume (e.g. a sphere of
 radius $R$), we have, for a given wave vector norm $k = |{\bf k}|$,
 about $(k R)^2$ adiabatic modes all
 characterized by the same two-mode density matrix.
 This is true before and after the reheating,
 and also irrespectively of the decoherence level.
 Finally this is still true when considering the projection
 of the adiabatic modes on the last scattering surface.
 Indeed, for sufficiently high angular momentum, 
 there exist an ensemble of well aligned two-modes
 with both members living on the last scattering
 surface \cite{CP2}. One can thus accumulate statistics to measure the
 four observables of Eq. \Eqref{Bell}.
 Unfortunately, an observational verification of the inequality Eq.
 \Eqref{Bell} seems excluded. Indeed the cosmic variance,
 which is of the order of the mean amplitude
 (hence proportional to $n \sim 10^{100}$),
 is much larger than the required precision, which is
 given by the spread of the coherent states ($=1$)
 \footnote{Could it be possible to verify, in principle,
 that a distribution is non-separable, i.e.  possesses quantum correlations
 and not only stochastic ones, without using the projectors of Eq.
 (\ref{proj}) ?
  This interesting question deserves further study.}

 The second question concerns the value of $\delta$
 in realistic inflationary models.
 This interesting question deserves further study, see the Added Note.
 Let us here simply compare the critical value $\delta=1$
 separating quantum from stochastic distributions to the expected level of
 non-Gaussianities.
 At the end of inflation, the two-point function of the
 gravitational potential $\Psi$ is conventionally
 \cite{Komatsu}
 parameterized by the coefficient $f_{N\!L}$
 entering the field redefinition
 $\Psi = \left[\phi + f_{N\!L} \, \phi^2/(a M)
 \right]/(\sqrt{\epsilon} M a)$
 where $\phi$ is our Gaussian field during inflation,
 $M$ is Planck mass, and $\epsilon$ the slow roll parameter.
 It has been observationally 
 limited to
 $-58 < f_{N\!L} < 134$ \cite{WMAP}, while theoretical calculations give
 $f_{N\!L} = O(10^{-2})$ for the inflationary phase.
 On one hand, the variation of the power spectrum of $\Psi$ is therefore
 $\Delta P/ P \simeq f_{N\!L}^2 P$
 where $P$ is the power spectrum
 in the linear approximation ($\simeq 10^{-10}$).
 On the other hand, using  \Eqref{momenta2},
 one gets $\Delta P/P = \delta /n$.
 Therefore $f_{N\!L}= 10^{-2}$ corresponds t
 $\delta \simeq n P f_{N\! L}^2 \simeq 10^{86}$.
 This indicates that the minimal source of decoherence, the
 non-linear interactions during inflation,
 should be strong enough to give rise to separable distributions.

Acknowledgements: We would like to thank Ulf Leonhardt and Serge
Massar for interesting discussions and suggestions.

\end{document}